\documentclass[12pt]{article}

\usepackage{graphics,graphicx} 
\graphicspath{{images/}} 

\usepackage[export]{adjustbox} 

\usepackage{fullpage,natbib,multirow}

\usepackage{amsmath,amssymb,verbatim,epsfig}
\usepackage[dvipsnames,usenames]{color}

\newtheorem{defin}{\bf Definition}

\def\ga{\mbox{Ga}}
\def\iga{\mbox{IGa}}

\def\be{\mbox{Be}}

\def\no{\mbox{N}}

\def\un{\mbox{Un}}

\def\E{\mbox{E}}
\def\V{\mbox{Var}}

\def\P{\mbox{Pr}}

\def\Cv{\mbox{Cov}}

\def\rest{\mbox{rest}}

\def\tr{\mbox{tr}}

\def\by{{\bf y}}
\def\bz{{\bf z}}

\def\bS{{\bf S}}

\def\bV{{\bf V}}

\def\bY{{\bf Y}}
\def\bZ{{\bf Z}}
\def\bzero{{\bf 0}}
\def\simind{\stackrel{\mbox{\scriptsize{ind}}}{\sim}}
\def\simiid{\stackrel{\mbox{\scriptsize{iid}}}{\sim}}

\newcommand{\bLambda}{\boldsymbol{\Lambda}}

\newcommand{\bmu}{\boldsymbol{\mu}}
\newcommand{\bnu}{\boldsymbol{\nu}}

\newcommand{\bSigma}{\boldsymbol{\Sigma}}
\newcommand{\bOmega}{\boldsymbol{\Omega}}

\newcommand{\PY}{\mathcal{PD}}

\newcommand{\I}{{\rm 1}\!\!{\rm 1}}

\begin{document}

\baselineskip=24pt

\title{ Model-based approach for household clustering with mixed scale
variables }
\author{{\sc Christian Carmona$^1$, Luis Nieto-Barajas$^2$ \& Antonio Canale$^3$} \\[2mm]
{\sl $^1$ Department of Statistics, University of Oxford} \\
{\sl $^2$ Department of Statistics, ITAM} \\
{\sl $^3$ Department of Statistical Sciences, University of Padua} \\[2mm]
{\small {\tt carmona@stats.ox.ac.uk, lnieto@itam.mx  {\rm and} canale@stat.unipd.it}} \\ }
\date{}
\maketitle

\begin{abstract}
The Ministry of Social Development in Mexico is in charge of creating and assigning social programmes targeting specific needs in the population for the improvement of quality of life. To better target the social programmes, the Ministry is aimed to find clusters of households with the same needs based on demographic characteristics as well as poverty conditions of the household. Available data consists of continuous, ordinal, and nominal variables and the observations are not iid but come from a survey sample based on a complex design. We propose a Bayesian nonparametric mixture model that jointly models a set of latent variables, as in an underlying variable response approach, associated to the observed mixed scale data and accommodates for the different sampling probabilities. The performance of the model is assessed via simulated data. A full analysis of socio-economic conditions in households in the Mexican State of Mexico is presented. 
\end{abstract}

\vspace{0.2in} \noindent {\sl Keywords}: Bayes nonparametrics, complex design, latent variables, multivariate normal, Poisson-Dirichlet process.

\section{Introduction}
\label{sec:intro}

The Ministry of Social Development (SEDESOL, according to its initials in Spanish) in Mexico is one of many government dependencies. The aim of SEDESOL is to help and improve the social backwardness that prevails in a high percentage of the households in the country. To fulfil this objective, SEDESOL creates social programmes to target specific needs in the population. 

Currently, each existing social programme has its own rules of operation and its own way of selecting the potential households to be benefited by the programme, but in general they all use household income as the main selection criterion. In order to simplify the selection of potential candidates and to better target the programmes to the correct population, SEDESOL wants to create a clustering of households based on needs, socio-economical and demographical features as well as poverty conditions. 

In 2009, the Council for National Evaluation \citep{coneval:09} proposed a methodology for measuring the poverty conditions in households in terms of multiple indicators. These include the income dimension, social deprivations and social cohesion. As a result of this new methodology a multi-dimensional measurement was created based on seven indicators: income, and six deprivation indicators such as education, access to health services, access to social security, housing quality, access to basic public services and access to feeding. These will be the core set of variables used in the clustering later on. 

In January of 2004 a new general law for social development was passed in Mexico. It establishes that poverty measurements must be calculated every two years at a state level and every five years at a municipality level. For these purposes, the National Institute for Official Statistics (INEGI) implemented a survey based on a complex design of households. This national survey of income and expenses in households (ENIGH) through a module of socio-economic conditions (MCS) collects the required information to produce the multi-dimensional poverty indicators at a household level and in some cases at an individual level. These household poverty indicators are then expanded with the corresponding sampling design weights to produce poverty indicators at a state level. 

Model-based clustering approaches \citep{mclachlan&basford:88,banfield&raftery:93} rely on a probability model-based on a finite or infinite mixture of sub-populations. Typically the sub-populations are assumed to be normal but other alternatives, to deal for example with possible asymmetric structures of the subpopulation, have also been considered \citep{rodr:walk:2014, cana:scar:2016}. In the case of mixed-scale data, \cite{everitt:88} introduced the use of latent variables and thresholding approach to deal with binary and ordinal variables. Although he did not perform an actual clustering, the author mentioned that his maximum likelihood estimation procedure can be potentially used for model fitting in clustering. More recently, \cite{fernandez&al:16} used ordinal data in a model-based clustering, and \cite{pledger&arnold:14} used mixtures in the correspondence analysis and scaling problems. Many authors have considered the use of Everitt's latent variable and thresholding approach to induce a multivariate mixed-scale density such as \cite{cai&al:11}, \citet{nore:pele:2012}, \cite{mcparland&al:14} or \citet{cana:duns:sii}. 

Particularly, \cite{mcparland&al:14} proposed a finite mixture model to accomplish a similar task as the one considered here. As in our motivating application, their setting consisted of mixed-scale data, and specifically of binary, ordinal and nominal variables. To accommodate the nominal variables, the authors proposed an extended version of the latent variables considered in \cite{everitt:88}. They further parametrized the mean of the latents via a factor model, reducing so the effective number of parameters to estimate but losing some information. Our approach is similar to \cite{mcparland&al:14} but differs in several ways: we include continuous variables to the mixed-mode set of variables, propose an infinite mixture, and do not reduce dimensionality in the parametrization of the mean of the latent variables. In addition, we account for the complex design setting of our data while the paper by \cite{mcparland&al:14} focuses only on iid data. 

Finite mixtures and infinite mixtures have different features, and of course, they are both valid models in practice. Infinite mixtures are appealing since there is no need to define an upper bound for the maximum number of groups in a clustering problem. If the data supports a small number of groups, an infinite mixture model is also able to detect them. Moreover, an infinite mixture model can also be defined a-priori to support a small/large number of groups in average by an appropriate specification of the hyper parameters. 

In summary, the aim of this work is to produce clusters of households based on the multidimensional poverty indicators considering that the available data come from a complex sampling design where each household has a different sampling probability. As mentioned above, the data is composed by observed mixed scale variables that include numeric (continuous and discrete) and categorical (ordinal and nominal) variables. We introduce a convenient set of latent variables associated to each of the observed ones and use a Bayesian nonparametric mixture of multivariate normals to model their latent multivariate density. The model also accounts for the different sampling probabilities of the selected households. We believe that no other model has been proposed that takes into account all these features. 

The layout of the remaining of the paper is as follows: we define the model and the priors used to induce the clustering in Section \ref{sec:model}. Section \ref{sec:post} contains posterior characterisation of the model parameters, with some implementation details included in Section \ref{sec:implement}. Simulation studies and the analysis of the motivating data set are presented in Section \ref{sec:data}. We end with some concluding remarks in Section \ref{sec:conclusion}.

\section{Model}
\label{sec:model}

\subsection{Observed and latent variables}
Consider an individual $i$ that is characterized by a multivariate response of dimension $p$, i.e., $\by_i=\{y_{ij},\;j=1\ldots,p\}$ and $i=1,\ldots,n$. Without loss of generality let us assume that the total number of variables $p$ is divided into $c$ continuous variables, $o$ ordinal variables, and $m$ nominal variables such that $p=c+o+m$. In summary, $\by_i'=(y_{i,1},\ldots,y_{i,c},y_{i,c+1},\ldots,y_{i,c+o},y_{i,c+o+1},\ldots,y_{i,c+o+m})$ for $i=1\ldots,n$. We note that numerical discrete with finite number of values are also allowed and can be treated as ordinal \citep{cana:duns:sii} and count variables, that is, numerical discrete with an infinite number of values can also be included similarly \citep{cana:duns:jasa}.

We associate to each response $\by_i$ of dimension $p$ a latent vector $\bz_i$ of dimension $q$ in the following way: 

\begin{itemize}
\item
For each \emph{continuous} variable $y_{ij}$, $j=1,\ldots,c$, we define a  transformed variable $z_{ij}=g_j(y_{ij})$, where $g_j(\cdot)$ is a normalising transformation, (possibly different) for each $j$, to  stabilize the variance and have a real support. 
\item
For each \emph{ordinal} variable $y_{ij}$, $j=c+1,\ldots,c+o$, that takes values in an ordered set $\{\vartheta_k\}$ with $K_j$ different values, we propose one latent $z_{ij}$ satisfying $y_{ij}=\vartheta_k$ iff $\gamma_{j,k-1}<z_{ij}\leq\gamma_{j,k}$, for $k=1,\ldots,K_j$ and $\{\gamma_{j,0},\ldots,\gamma_{j,K_j}\}$ are fixed thresholds or cut-off points with $\gamma_{j,0}=-\infty$ and $\gamma_{j,K_j}=\infty$. More details on the choice of the cut-offs are discussed in the next Section. Note that a binary variable is a special case of an ordinal variable with $K_j=2$.
\item
For each \emph{nominal} variable $y_{ij}$, $j=c+o+1,\ldots,c+o+m$, with $L_j$ categories we propose to define $L_j-1$ latent variables that can be placed in the latent vector $\bz_{ij}=\{z_{il},\;l=c+o+\sum_{h=c+o+1}^{j-1}(L_h-1)+1,\ldots,c+o+\sum_{h=c+o+1}^{j}(L_h-1)\}$ such that 
$$y_{ij}=\left\{
\begin{array}{ll} 
L_j, & \mbox{if}\;\max_l(z_{i,l})<0 \\
k, & \mbox{if}\;z_{i,s}=\max_l(z_{i,l})\;\&\;z_{i,s}>0 \\
 & \mbox{with}\; s=c+o+\sum_{h=c+o+1}^{j-1}(L_h-1)+k,\;\mbox{and}\;k=1,\ldots,L_j-1.
\end{array}
\right.$$
\end{itemize}
Thus, latent vector $\bz_i$ is $q-$dimensional where $q=c+o+\sum_{j=c+o+1}^{c+o+m}(L_j-1)$. In summary $\bz_i'=(z_{i,1},\ldots,z_{i,c},z_{i,c+1},\ldots,z_{i,c+o},\bz_{i,c+o+1},$ $\ldots,\bz_{i,c+o+m})$ which for $j=c+o+1,\ldots,c+o+m$ each $\bz_{i,j}$ is of dimension $L_j-1$. Note that the $\{z_{ij}\}$ variables associated to the continuous variables are not actually latent, but a transformation of the observed $y_{ij}$. Strictly speaking this transformation is not needed since a Bayesian nonparametric mixture of normals has full support \citep{lo:84}.

Available data come from a complex survey sample where each individual $\by_i$ has known sampling probability $\pi_i$, $i=1,\ldots,n$. The reciprocal of these sampling probabilities, $w_i=1/\pi_i$, are called sampling design weights or expansion factors. 

There are different strategies to include the sample probabilities into an inferential procedure. In linear regression, \cite{lumley:10} proposes a weighted least squares with weights defined by $w_i$, that is $\min\sum_{i=1}^n \frac{1}{\pi_i}\left(y_i-\alpha-\beta x_i\right)^2$, where $x_i$ is an explanatory variable, $\alpha$ an intercept and $\beta$ the regression coefficient. \cite{chambers&skinner:03}, on the other hand, propose to re-weight the likelihood contribution with an exponent given by the sampling weight $w_i$, i.e., $\prod_{i=1}^n f(y_i\mid\theta)^{1/\pi_i}$, where $\theta$ is the model parameter. Considering Lumley's approach under a normal model, the weighted least squares estimates are equivalent to the maximum likelihood estimators with variances scaled by a factor $\pi_i$. Moreover, under a normal model Chambers and Skinner's approach is also equivalent to scale the variance with a factor $\pi_i$. 

To convince ourselves that scaling the variance in a normal model is a reasonable way to account for the sampling weights, consider $y_1,\ldots,y_w$ to be a sample of size $w$ from a $\no(\mu,\sigma^2)$ model, and assume that $\sigma^2$ is know. Then, the sufficient statistic for $\mu$ is $\bar{y}$ whose sampling distribution is $\bar{y}\sim\no(\mu,\sigma^2/w)$. So, if $\bar{y}$ is the only value we observe that represents $w$ individuals, its likelihood contribution would be to scale the variance of one individual, $\sigma^2$, by a factor $1/w=\pi$. 

Therefore we propose the following Gaussian weighted model for the latent vector:
\begin{equation}
\label{eq:model}
\bz_i\mid\bmu_i,\bSigma\sim\no_q(\bmu_i,\kappa\,\pi_i\bSigma),
\end{equation}
where $\no_q$ stands for a multivariate normal distribution of dimension $q$, $\bmu_i$ is a mean vector of dimension $q\times 1$, $\bSigma$ is a variance-covariance matrix of dimension $q\times q$ and $\kappa>0$ is a scaling factor that controls the impact of the sampling probabilities in the variance. This latter parameter will play a central role in the posterior clustering structures as discussed later in Section~\ref{sec:data}. 

Since there is a deterministic relationship between $\by_i$ and $\bz_i$, and some of the variables in $\by_i$ are categorical (ordinal or nominal), some constrains have to be imposed in the matrix $\bSigma$ for estimation purposes. For both types of categorical variables, the corresponding latents are partially determined by the observed data, so only its mean can be estimated and their variance have to be kept fixed. If we denote by $\sigma_j^2$ the variance of the $j$th variable and by $\sigma_{j,k}$ the covariance between variables $j$ and $k$ for $j,k=1,\ldots,n$, then: 
\begin{enumerate}
\item[(i)] 
For continuous variables, i.e. $j=1,\ldots,c$ : $\sigma_j^2>0$, 
\item[(ii)] 
For ordinal variables, i.e. $j=c+1,\ldots,c+o$ : $\sigma_j^2=1$, 
\item[(iii)] 
For nominal variables, i.e. $j=c+o+1,\ldots,\sum_{h=c+o+1}^{c+o+m}(L_h-1)$ : $\sigma_j^2=1$.
\end{enumerate}
Additionally, $\sigma_{jk}$'s are such that the implied variance-covariance matrix $\bSigma$ is positive defined. 

\subsection{Prior distributions}

The clustering model will be based on an appropriate choice of the prior distribution on the $\{\bmu_i\}$. A clustering of the $\{\bmu_i\}$ will induce a clustering of the $\{\by_i\}$. For that we require that the marginal distribution for each $\bmu_i$ to have a continuous support on the real line and the joint distribution to satisfy that $\P(\bmu_i=\bmu_{i'})>0$ for all $i\neq i'\in\{1,\ldots,n\}$. There are several nonparametric priors that satisfy these conditions. A sufficiently rich class is the (two-parameters) Poisson-Dirichlet process \citep{pitman&yor:97} that includes the Dirichlet \citep{ferguson:73} and the normalized stable \citep{kingman:75} processes as particular cases. 

A Poisson-Dirichlet process $G$ is almost surely discrete \citep[see e.g.][]{ishwaran&james:01} and is defined as follows: 
$$G(\cdot)=\sum_{k=1}^\infty \omega_k \delta_{\xi_k}(\cdot),$$
where $\delta_{\xi}$ denotes a point mass at $\xi$ and $\{\omega_k\}$ are the weights. Here, both the weights $\omega_k$ and the locations $\xi_k$ are random variables such that $\xi_k\simiid G_0$ for all $k=1,2,\dots$, with $G_0$ a probability distribution. The weights $\{\omega_{k}\}$ are defined as $\omega_1=\nu_1$ and $\omega_k=\nu_k\prod_{l<k}(1-\nu_l)$, with $\nu_k\simind\be(1-a,b+ka)$, where $\be(a,b)$ stands for a beta distribution with mean $a/(a+b)$. This specific choice for the distribution of the stick-breaks $\nu_k$ characterizes the Poisson-Dirichlet process. The functional parameter $G_0$ is known as centering measure since $\E(G)=G_0$. In notation we say $G\sim\PY(a,b,G_0)$, where $\PY$ stands for a Poisson-Dirichlet process with parameters $a\in[0,1)$, $b>-a$ and centering measure $G_0$. 
The Dirichlet and the normalized stable processes arise when $a=0$ and $b=0$, respectively. 

Therefore our prior on the $\{\bmu_i\}$ will be 
\begin{equation}
\label{eq:priormu}
\bmu_i|G\simiid G,\;\;\mbox{for }\;i=1,\ldots,n\;\mbox{ with }\; G\sim\PY(a,b,G_0),
\end{equation}
and $G_0(\bmu)=\no(\bzero,\bSigma_\mu)$, where $\bSigma_\mu=\mbox{diag}(\sigma^2_{\mu_1},\ldots,\sigma^2_{\mu_q})$. In consequence, this choice of prior implies that the $\{\bmu_i\}$ are exchangeable with marginal distribution $\bmu_i\sim G_0$ for all $i=1,\ldots,n$. \cite{pitman:95} showed that if we integrate out the nonparametric measure $G$, the joint distribution of the $\{\bmu_i\}$ is characterized by a generalized Polya urn mechanism with conditional distribution that depends on the density $g_0$ associated to $G_0$ and given by
\begin{equation}
\label{eq:purnc}
f(\bmu_i\mid \bmu_{-i})=\frac{b+a\,r_i}{b+n-1}g_0(\bmu_i)+\sum_{j=1}^{r_i}\frac{n_{j,i}^*-a}{b+n-1}\delta_{\bmu_{j,i}^*}(\bmu_i),
\end{equation}
for $i=1,\ldots,n$, where $\bmu_{-i}=(\bmu_1,\ldots,\bmu_{i-1},\bmu_{i+1},\ldots,\bmu_n)$ denotes the set of all $\{\bmu_j\}$ excluding the $i^{th}$, and $(\bmu_{1,i}^*,\ldots,\bmu_{r_i,i}^*)$ denote the unique values in $\bmu_{-i}$, each occurring with frequency $n_{j,i}^*$, $j=1,\ldots,r_i$, which satisfy the condition $n_{1,i}^*+\cdots+n_{r_i,i}^*=n-1$. Therefore $\P(\bmu_i=\bmu_{i'})>0$ as desired. The number of unique values $r$ in $\bmu=(\bmu_1,\ldots,\bmu_n)$ determines the number of clusters. This value is controlled by the parameters $(a,b)$. Larger values of either $a$ or $b$, within the valid ranges, produce a larger $r$ \cite[e.g.][]{navarrete&al:08}. It is worth noting that in our Poisson-Dirichlet mixture model the $\{\bmu_i\}$ are unobserved parameters, so the clustering structure is more complex than that directly induced by random mixing measure. Indeed it is also controlled by the Gaussian kernel and by its scale parameters. In Section \ref{sec:sim1} we illustrate this point through a simulation study. 

The mixture specification allows the $\gamma_{jk}$ to be fixed as discussed also in \citet{kottas}, \citet{leon&al:10}, and \citet{cana:dipa}. 
Note that without the mixture specification, i.e. assuming a single multivariate normal density, the model is a multivariate ordered probit model where for the sake of identifiability, only each first marginal cut-off is fixed.
Our nonparametric mixture specification, instead, provides enough flexibility to fix all the internal cut-offs. Different probability masses for different levels of the ordered categories are obtained placing different kernel densities within each interval. Specifically we fix the internal cut-offs at $\gamma_{j,k}\in\{0,\pm 4, \pm 8, \ldots\}$. This choice along with the fixed variance specification (ii), force the probability masses of each internal category to be at most $0.95$ for $k=2,\dots, K_j-1$, keeping the first and last cell probabilities unrestricted. This assumption is not restrictive for our motivating application and has the advantage of simplifying posterior inference while improving the mixing of the Markov Chain Monte Carlo (MCMC) sampling. If needed, however, the fixed variance specification (ii) may be be removed while keeping the fixed cut-offs. 

Considering the variance-covariance matrix $\bSigma$ we recall that it has several constraints on the diagonal given by conditions (i)--(iii). To include them and to ensure positive definiteness, we follow a separation strategy as in \cite{barnard&al:00} such that $\bSigma=\bLambda\bOmega\bLambda$ where $\bLambda=\mbox{diag}(\sigma_1,\ldots,\sigma_q)$ is a diagonal matrix of standard deviations and $\bOmega$ is a correlation matrix. We thus assign independent priors on the squared elements of $\bLambda$ and on $\bOmega$. 

For those $j=1,\ldots,q$ such that $\sigma_j^2\neq 1$ we take 
\begin{equation}
\label{eq:priorvar}
\sigma_j^2\simiid\iga(d_0^z,d_1^z),
\end{equation}
where $d_0^z,d_1^z>0$, and for the correlation matrix we take 
\begin{equation}
\label{eq:priorcor}
f(\bOmega)\propto|\bOmega|^{q(q-1)/2-1}\left(\prod_{j}|\bOmega_{jj}|\right)^{-(q+1)/2},
\end{equation}
where $\bOmega_{jj}$ is the $i$th principal sub-matrix of $\bOmega$.
This prior implies that individual correlations are marginally uniform on the interval $[-1,1]$ \citep{barnard&al:00}.

We finally take hyper-priors for the parameters $a$, $b$ and $\sigma_{\mu j}$ as 
\begin{eqnarray}
\nonumber
f(a)=\alpha \delta_{0}+(1-\alpha)\be(a\mid d_0^a,d_1^a),\;\;\;f(b\mid a)=\ga(b+a\mid d_0^b,d_1^b), \\
\label{eq:hpriors}
\mbox{and}\;\;\;\sigma_{\mu j}^2\simiid\iga(d_0^\mu,d_1^\mu)\;\; j=1,\ldots,q, \hspace{2.5cm}
\end{eqnarray}
where $\ga(a,b)$ stands for a gamma distribution with mean $a/b$, $\iga(a,b)$ stands for an inverse gamma distribution with mean $b/(a-1)$, and the values $d_0^a,d_1^a,d_0^b,d_1^b,d_0^{\mu},d_1^{\mu}$ are all positive constants. 
Note that the hyper-prior on $a$ is a mixture of a point mass at zero and a continuous beta distribution. This is to consider the option that $a=0$ with positive probability. The hyper prior on $b$ is given conditionally on $a$ and includes the constraint that $b>-a$ by shifting the support of the gamma density to the interval $(-a,\infty)$.

\section{Posterior characterisation}
\label{sec:post}

Let $(\bmu,\bSigma,a,b,\bSigma_{\mu})$ the set of parameters and hyper-parameters of the model, where $\bmu'=(\bmu_1,\ldots,\bmu_n)$. The posterior distribution of these parameters is characterised in terms of its full conditional distributions which are given below. In what follows ``$\rest$'' means all other parameters and the data.
\begin{enumerate}
\item[(a)] The conditional posterior distribution of $\bmu_i$, $i=1,\ldots,n$, is 
$$f(\bmu_i\mid\bz,\bmu_{-i})=p_0\no_q(\bmu_i\mid \bnu_i,\bV_i)+\sum_{j=1}^{r_i}p_j\delta_{\bmu_{j,i}^*}(\bmu_i),$$
where $\bnu_i=\bV_i(\pi_i\kappa\bSigma)^{-1}\bz_i$, $\bV_i=\left((\pi_i\kappa\bSigma)^{-1}+\bSigma_{\mu}^{-1}\right)^{-1}$, $p_j=D_j/(\sum_{l=0}^{r_i}D_l)$ for $j=0,\ldots,r_i$ with $D_0=(b+ar_i)\no_q(\bz_i\mid\bzero,\pi_i\kappa\bSigma+\Sigma_{\mu})$ and, for $j>0$, $D_j=(n_{ji}^*-a)\no_q(\bz_i\mid\bmu_{ji}^*,\pi_i\kappa\bSigma)$. 
\end{enumerate}

Conditional distribution (a) allows us to identify which $\mu_i$'s are equal to each other or different. We further need to re-sample the unique values $(\bmu_i^*,\ldots,\bmu_r^*)$ in $(\bmu_1,\ldots,\bmu_n)$. 
\begin{enumerate}
\item[(b)]
Conditional on the membership allocations $I_j=\{i:\,\bmu_i=\bmu_j^*\}$, $j=1,\ldots,r$
$$f(\bmu_j^*\mid\bz,I_j,\rest)=\no(\bmu_j^*\mid\bnu_j^*,\bV_j^*),$$
where $\bnu_j^*=\frac{1}{\kappa}\bV_j^*\bSigma^{-1}\left(\sum_{i\in I_j}(1/\pi_i)\bz_i\right)$ and $\bV_j^*=\left(\left(\frac{1}{\kappa}\sum_{i\in I_j}(1/\pi_i)\right)\bSigma^{-1}+\bSigma_{\mu}^{-1}\right)^{-1}$. 
\end{enumerate}
\begin{enumerate}
\item[(c)]
The conditional posterior distribution of the diagonal elements $\sigma_{\mu l}^2$, $l=1,\ldots,q$, of matrix $\bSigma_\mu$ is
$$f\left(\sigma_{\mu l}^2\mid\rest\right)=\iga\left(\sigma_{\mu l}^2\left|\,d_0^{\mu}+\frac{r}{2},\,d_1^{\mu}+\frac{1}{2}\sum_{j=1}^r\left(\mu_{jl}^*\right)^2\right.\right),$$
where $\mu_{jl}^*$ is the $l$-th coordinate of vector $\bmu_j^*$. 
\end{enumerate}

For the matrix $\bSigma$, considering the separation strategy, the full conditional distributions of $\bLambda$ and $\bOmega$ are 
\begin{enumerate}
\item[(d)] The conditional posterior distribution of the squared diagonal elements of $\bLambda$, $\sigma_{j}^2$, for those $j$ such that $\sigma_{j}^2\neq 1$, is given by
$$f(\sigma_{j}^2\mid\bz,\rest)\propto (\sigma_j^2)^{-(d_0^z+n/2+1)} e^{-d_1^z/\sigma_j^2} \exp\left\{-\frac{1}{2}\tr(\bSigma^{-1}\bS)\right\},$$
where $\bS=\sum_{i=1}^n\frac{1}{\kappa\pi_i}(\bz_i-\bmu_i)(\bz_i-\bmu_i)'$.
\item[(e)] The conditional posterior distribution of $\bOmega$ is simply
$$f(\bOmega\mid\bz,\bLambda,\rest)\propto\left(\prod_{j=1}^q|\bOmega_{jj}|\right)^{-\frac{1}{2}(q+1)}\left|\bOmega\right|^{-\frac{1}{2}\{n+2-q(q-1)\}}\exp\left[-\frac{1}{2}\mbox{tr}\left\{\bOmega^{-1}\left(\bLambda^{-1}\bS\bLambda^{-1}\right)\right\}\right],$$
where $\bS$ is given above.
\end{enumerate}

The likelihood for $a$ and $b$ is given by the exchangeable partition probability function induced by the Poisson-Dirichlet process \citep{pitman:95}. Thus,
\begin{enumerate}
\item[(f)] The conditional posterior distribution of $a$ is
$$f(a\mid b,\rest)\propto \left\{\prod_{j=1}^{r-1}(b+ja)\right\}\left\{\prod_{j=1}^r\frac{\Gamma(n_j^*-a)}{\Gamma(1-a)}\right\}f(a),$$
where the prior $f(a)$ is given in \eqref{eq:hpriors}.
\item[(g)] The conditional posterior distribution of $b$ is
$$f(b\mid a,\rest)\propto \frac{\Gamma(b+1)}{\Gamma(b+n)}\left\{\prod_{j=1}^{r-1}(b+ja)\right\}f(b\mid a),$$
where $f(b\mid a)$ is also given in \eqref{eq:hpriors}.
\end{enumerate}

Finally, the latent variables $z_{ij}$ have to be re-sampled from their respective predictive distributions following the constraints mentioned at the beginning of Section \ref{sec:model}. 
\begin{enumerate}
\item[(h)] The conditional predictive distributions for $z_{ij}$, $i=1,\ldots,n$ are: 
\begin{itemize}
\item For $j=1,\ldots,c$, $z_{ij}$ remains unchanged.
\item For $j=c+1,\ldots,c+o$, $z_{ij}$ has a truncated normal distribution of the form $$z_{ij}\mid \bz_{-(ij)},\rest\sim\no(\nu_{ij},V_{ij})I(\gamma_{j,k-1}<z_{ij}\leq\gamma_{j,k}),$$
for $y_{ij}=k$, where
$\nu_{ij}=\mu_{ij}+\bSigma_{12}\bSigma_{22}^{-1}\left(\bz_{-(ij)}-\bmu_{-(ij)}\right)$ , $V_{ij}=\bSigma_{11}-\bSigma_{12}\bSigma_{22}^{-1}\bSigma_{21}$, with $\bSigma_{11}=\V(z_{ij})=\sigma_j^2$, $\bSigma_{12}=\Cv(z_{ij},\bz_{-(ij)})$, and $\bSigma_{22}=\V(\bz_{-(ij)})$. 
\item For $j=c+o+1,\ldots,c+o+\sum_{h=c+o+1}^m(L_h-1)$, the latent vector $\bz_{ij}=\{z_{il}\}$, $l=c+o+\sum_{h=c+o+1}^{j-1}(L_h-1),\ldots,c+o+\sum_{h=c+o+1}^j(L_h-1)$ has $z_{il}$ element coming from a truncated normal distribution of the form
$$z_{il}\mid\bz_{il},\rest\sim\no(\nu_{il},V_{il})I_{A_{ij}}(z_{il}),$$
where $$A_{ij}=\left\{\begin{array}{ll}
(-\infty,0) & \mbox{if}\;y_{ij}=L_j \\
\left(-\infty,z_{is}\right) & \mbox{if}\;y_{ij}=k<L_j\;\&\;l\neq s=c+o+\sum_{h=c+o+1}^{j-1}(L_h-1)+k \\
\left(\max_s\{z_{is},0\},\infty\right) & \mbox{if}\;y_{ij}=k<L_j\;\&\;s\neq l=c+o+\sum_{h=c+o+1}^{j-1}(L_h-1)+k
\end{array}\right.$$
and $\nu_{il}$ and $V_{il}$ are as above. 
\end{itemize}
\end{enumerate}

\section{Implementation details and clustering selection}
\label{sec:implement}

Posterior inference of our model will rely on the implementation of a MCMC procedure. 
In what follows, $\varphi_{\sigma}$, $\varphi_{\rho}$ and $\varphi_{b}$ are tuning parameters that control de acceptance probability of the corresponding Metropolis-Hastings (MH) steps \citep{tierney:94}. For the examples considered here we took $\varphi_{\sigma}=5$, $\varphi_{\rho}=4$ and $\varphi_{b}=2$ to achieve reasonable acceptance probabilities. 

Sampling from conditional distributions (a)--(c) is straightforward. To sample from conditional distribution (d) we propose a random walk MH. At iteration $(r+1)$ we sample from the proposal distribution ${\sigma_{j}^2}^*\sim\ga\left(\varphi_{\sigma},\varphi_{\sigma}/{{\sigma_{j}^2}^{(r)}}\right),$ with $\varphi_\sigma>0$. 
We first check positive definiteness of $\bSigma^*$ with the new draw and accept with its corresponding probability. 

For conditional distribution (e) we proceed to sample conditionally one at a time the elements $\rho_{jk}$ of $\bOmega$. Using \cite{barnard&al:00}'s ideas, we obtain the support of $\rho_{jk}$ that keeps $\bOmega$ positive defined by computing the roots of the quadratic function $h(\rho)=\theta_1\rho^2+\theta_2\rho+\theta_3$, where $\theta_1=\{h(1)+h(-1)-2h(0)\}/2$, $\theta_2=\{h(1)-h(-1)\}/2$ and $\theta_3=h(0)$ with $h(\rho)=|\bOmega(\rho)|$ and $\bOmega(\rho)$ is the correlation matrix $\bOmega$ evaluated in $\rho$ at the entry $j,k$. 
We can then use the griddy Gibbs sampler \citep{ritter&tanner:92} and evaluate the conditional density top sample from $\rho_{jk}$. However, from our experience, this procedure highly increases the computational time. Instead we implement a random walk MH step with uniform proposal distribution. If we denote by $[\rho_1,\rho_2]$ the support of $\rho_{jk}$ and by $\ell=\rho_2-\rho_1$ its length, the proposal is taken as $\rho_{jk}^*\sim\un\left(\max(\rho_1,\rho_{jk}^{(r)}-\ell/\varphi_{\rho}),\min(\rho_2,\rho_{jk}^{(r)}+\ell/\varphi_{\rho})\right)$, with $\varphi_\rho>0$. 

To sample from conditional distributions (f) we implement a MH step with  independent proposal distribution $p(a)=1/2\delta_{0}(a) + 1/2\be(1,1)$. For (g) we propose a random walk MH step with proposal distribution as the one used for $\rho_{jk}$, i.e $b^*\mid b^{(r)}\sim\un(b^{(r)}-\varphi_{b},b^{(r)}+\varphi_{b})$, for $\varphi_b>0$, and always considering the support $b>-a$. 

Summarising the posterior distribution assigned to all possible partitions of the data is not an easy task. This posterior distribution is characterised through a MCMC sample. At each iteration of the MCMC sampler an $n\times n$ adjacency matrix containing a 1 in position $ij$ if elements $i$ and $j$ share the same value of $\bmu^*$ is stored. At the end of the Gibbs sampler, a similarity matrix is computed as the Montecarlo average of all the adjacency matrices. This similarity matrix represents the ``average clustering''. As posterior summary of all clustering structures available, we select the adjacency matrix of the iteration with minimum squared distance from the average similarity matrix. This procedure was originally proposed by \cite{dahl:06} and has been recently formalised in a decision theory framework by \cite{wade&ghahramani:15}, so it becomes an optimal decision for a specific loss function. Alternatively, other decision criteria based on MAP (maximum a-posteriori probabilities) could be used.

This posterior inference procedure and the clustering selection has been implemented in {\sf R} \citep{r:16} in the package {\tt BNPMIXcluster} \citep{carmona&nieto:17} that is available from The Comprehensive R Archive Network (CRAN). 

Since the observed data is of mixed mode, it is not easy to define a clustering comparison measure to compare among the clusters obtained by different prior specifications. One possibility would be to use the point predictors of the underlying latent vectors, say $\hat{z}_{ij}$ obtained as the posterior predictive mean. However, the different prior specifications induce totally different values for the latent variables up to the point that they are not comparable. Therefore we only rely on the observed data to define such a measure. To try to get rid of the scales of the variables, we define new variables $y_{ij}^*$ as: for a numerical variable $j$ (continuous or discrete) $y_{ij}$ is standardized across all individuals $i=1,\ldots,n$; for a categorical variable $j$, if the number of categories is two then $y_{ij}^*=y_{ij}$, otherwise define $y_{il}^*$ a latent indicator variable for each category $l=1,\ldots,L_j$. Now, if $C_1,\ldots,C_r$ denote the $r$ groups associated to a particular clustering with group sizes $n_1,\ldots,n_r$ respectively, following \cite{nieto&contreras:14}, we summarize the heterogeneity of a clustering by a heterogeneity measure (HM) based on weighted variances in the following way 
\begin{equation}
\label{eq:hm}
\mbox{HM}(C_1,\ldots,C_r)=\sum_{k=1}^r n_k \sum_{j=1}^{p^*}S_{kj}^2,\quad
\mbox{where}\quad S_{kj}^2=\sum_{i=1}^{n_k} w_i^{(k)}{y_{ij}^*}^2-\left\{\sum_{i=1}^{n_k} w_i^{(k)}y_{ij}^*\right\}^2,
\end{equation}
with $w_i^{(k)}=w_i/\{\sum_{l\in C_k} w_l\}$ the normalized weights of individuals in cluster $k$ and $p^*$ is the number of resulting $\{y_{ij}^*\}$ variables. The larger the value of HM the more heterogeneous a clustering is. As acknowledged by \cite{nieto&contreras:14} these values should
be compared with care across different clusterings since in the extreme case that each individual forms its own cluster then HM takes the value of zero. So it is preferably a clustering with small HM and small $r$.

\section{Data analyses}
\label{sec:data}

\subsection{Simulation study 1}
\label{sec:sim1}

\begin{figure}
\centerline{\includegraphics[scale=0.8]{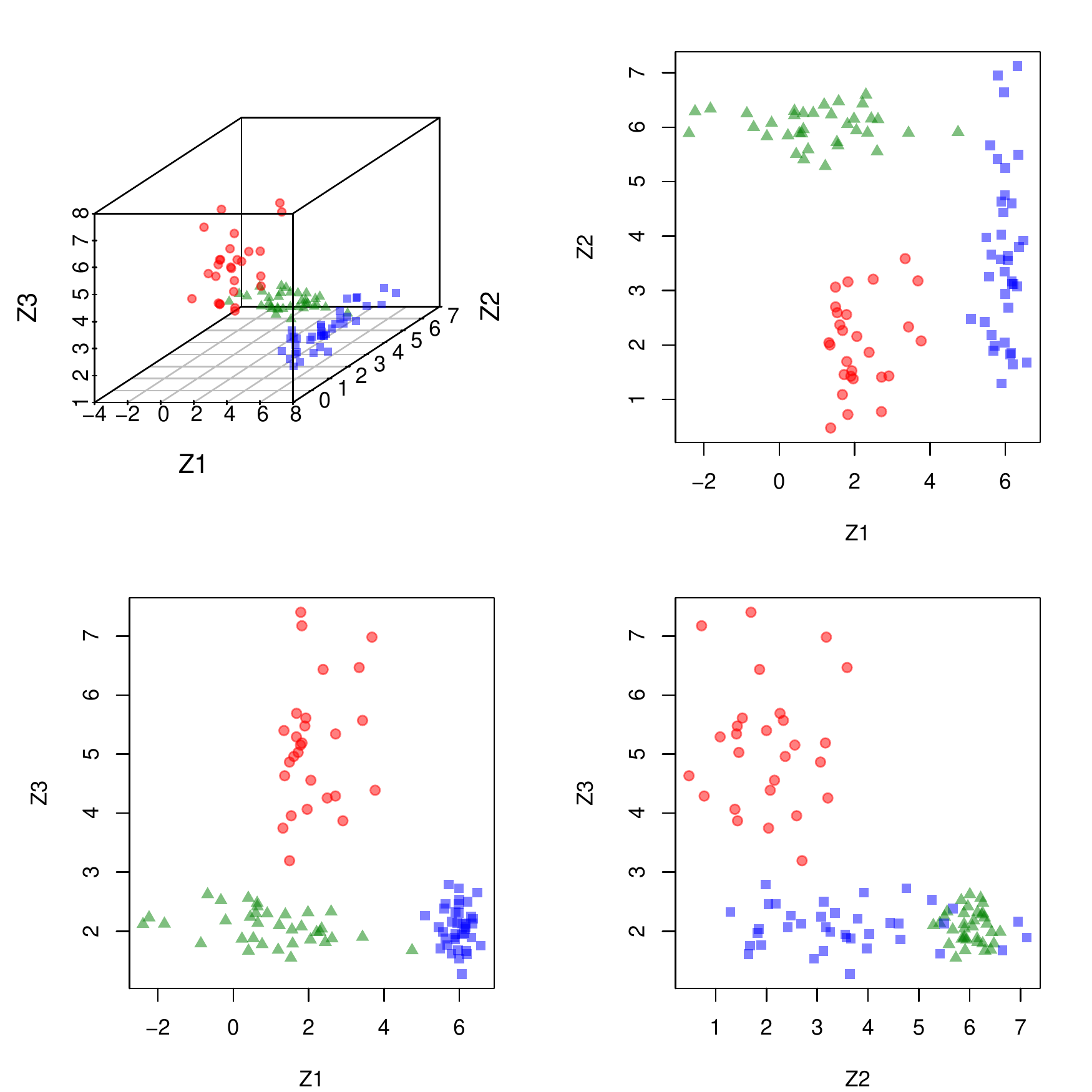}}
\caption{Simulated data set over the three latent variables $Z_1,Z_2,Z_3$. Each symbol represents a group: green triangles, blue squares and red circles.}
\label{fig:simdata}
\end{figure}

We first evaluate the clustering performance of our model in the presence/absence of categorical variables with independent and identically distributed data. 
For that, we sampled 3-dimensional latent continuous vectors $\bz'=(z_1,z_2,z_3)$ from a 3-components mixture of normals with equal mixing probabilities and mixture components with means $\bmu_1'=(2,2,5)$, $\bmu_2'=(6,4,2)$ and $\bmu_3'=(1,6,2)$, and variances $\bSigma_1=\mbox{diag}(1,1,1)$, $\bSigma_2=\mbox{diag}(0.1,2,0.1)$ and $\bSigma_3=\mbox{diag}(2,0.1,0.1)$. A sample from this model is included in Figure \ref{fig:simdata}. The data clearly show the existence of three groups.  

We tested this model by considering different scenarios for the observed data that combine the use continuous and discretised versions of the simulated data. These scenarios are: 
\begin{enumerate}
\item[(I)] Three continuous variables $(y_1,y_2,y_3)$ defined as $y_i=z_i$, for $i=1,2,3$.
\item[(II)] Two binary variables $(y_1,y_3)$ defined as $y_1=\I(z_1>5)$ and $y_3=\I(z_3>3)$.
\item[(III)] Two binary variables $(y_1,y_3)$ defined as in Scenario (II), one ordinal variable $y_2$ such that $y_2=\I(4<z_2\leq 5)+2\I(z_2>5)$, and a continuous variable  $y_4\sim\no(0,1)$.
\end{enumerate}
The function $\I(A)$ denotes the indicator function that takes the value one if the condition $A$ is satisfied and zero otherwise. 

Scenario I takes the latent continuous vectors as the observed data, so it is expected that the model would not have any problem in detecting the true clustering structure. Scenario II, however, considers only two binary variables defined in terms of the latents $z_1$ and $z_3$, respectively. If we look at Figure \ref{fig:simdata} we note that binary variables $I(z_1>5)$ and $I(z_3>3)$ are enough to distinguish the three groups, so the idea with this Scenario II is to test the definition of the continuous underlying latent variables through thresholding. Finally, Scenario III is the most challenging one because it adds to Scenario II two more variables: one ordinal, $y_2$ based on $z_2$; and one continuous, $y_4$ taken from a standard normal. These two additional variables act as noisy variables since the clustering is already defined by the binary variables $y_1$ and $y_3$. 

To implement our model, the cut-off points were defined as mentioned in Section \ref{sec:model}, that is, $(\gamma_0,\gamma_1,\gamma_2)=(-\infty,0,\infty)$ for the binary variables, and $(\gamma_0,\gamma_1,\gamma_2,\gamma_3)=(-\infty,0,4,\infty)$ for the ordinal variable with 3 categories. 

The priors for the parameters $a$ and $b$ that define the Poisson-Dirichlet process have hyper-parameters: $\alpha=0.5$, $d_0^a=d_1^a=d_0^b=d_1^b=1$. In all cases we took $\kappa=1$ and $\pi_i=1$. It is well-known that the clustering properties of the Bayesian nonparametric mixture models highly rely on the variances \citep[e.g.][]{barrios&al:13,nieto&contreras:14}. Therefore, we assess the performance of the model by considering different choices for the parameters of the priors on the variances $\sigma_j^2$ and $\sigma_{\mu j}^2$. These choices are: 
\begin{enumerate}
\item[A)] $d_0^z=d_0^{\mu}=0.1$ and $d_1^z=d_1^{\mu}=0.1$
\item[B)] $d_0^z=d_0^{\mu}=1$ and $d_1^z=d_1^{\mu}=1$
\item[C)] $d_0^z=d_0^{\mu}=2.1$ and $d_1^z=d_1^{\mu}=30$
\end{enumerate}

Both prior specifications (A) and (B) are vague and induce a prior variance that goes to infinity. Between the two, prior (A) is the most vague, almost non-informative.
On the other hand, specification (C) is slightly informative so to induce a prior mean of 27 and a prior variance of 7438. Although the latter specification is slightly more informative that the first two, it can also be considered vague since its variance is still very large. 

Combining Scenarios I--III and prior specifications (A)--(C) we have a total of 9 runs. For each of them a sample of size $n=100$ was taken and a Gibbs sampler with 4700 iterations was implemented with a burn-in of 200 and a thinning of 3. A total of 1500 MCMC draws were kept for inference. We ran our model in a Unix processor @ 1.60 GHz with 4 cores and 16 GB of RAM. For each of the priors (A)--(C), Scenario I took 71, 49, and 19 minutes, respectively; Scenario II 40, 40, and 36 minutes and  Scenario III 85, 78, and 53 minutes.

\begin{figure}
\centerline{\includegraphics[scale=0.5]{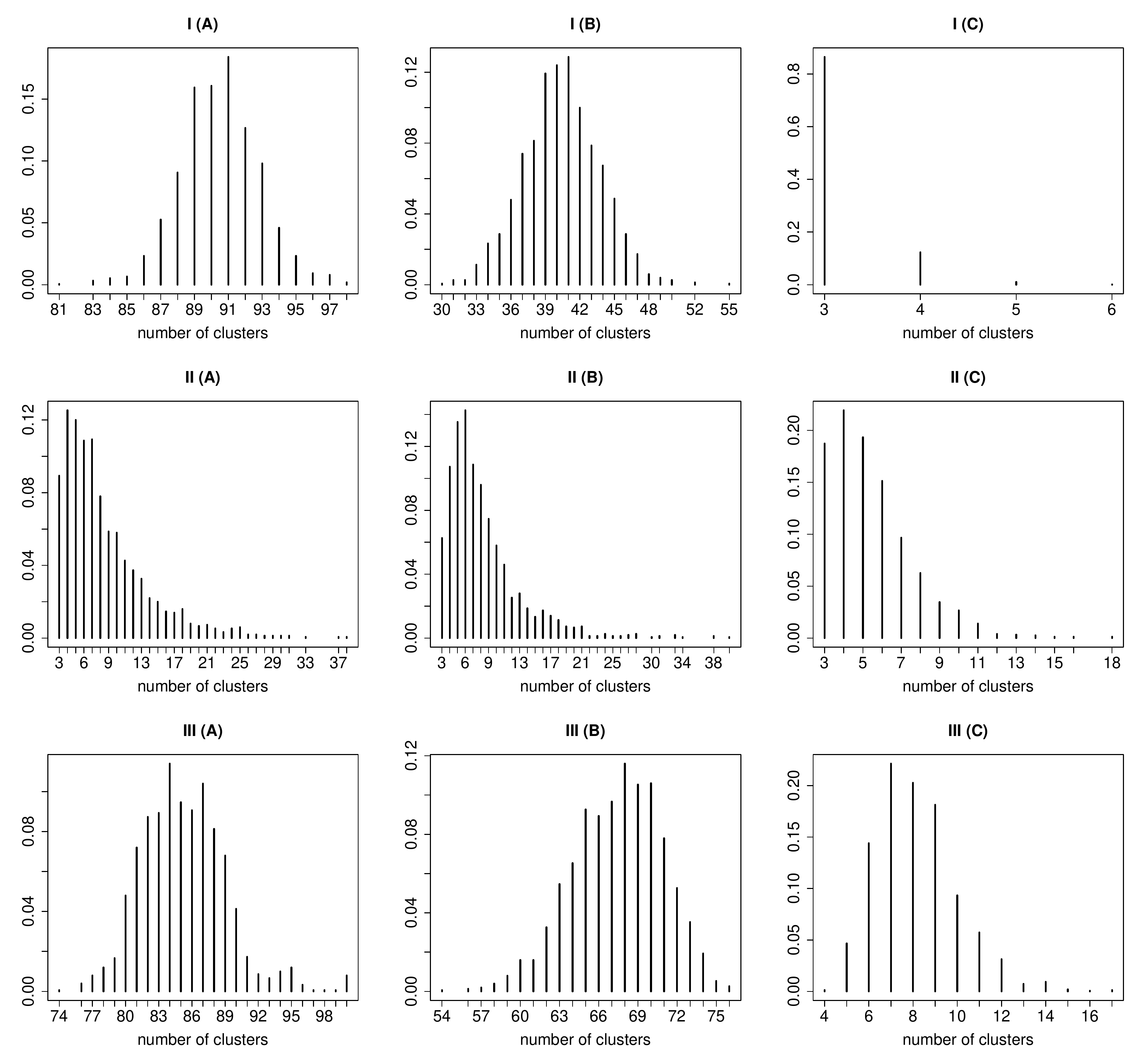}}
\caption{Probability histograms of the number of clusters for scenarios I, II and III (in the rows) and prior specifications (A), (B) and (C) in the columns, as described in Section \ref{sec:sim1}.}
\label{fig:nclust}
\end{figure}

We summarised the clustering performance by computing the posterior probability of the number of groups obtained by the different runs. Results are reported in Figure \ref{fig:nclust}. As we can see the choice of the prior parameters on the variances is crucial for the performance of the model. The number of clusters reduces when we go from prior (A) to (C) (columns one to three in Figure~\ref{fig:nclust}, respectively). The reduction is more drastic in the presence of continuous variables like in Scenarios I (first row) and Scenario III (third row). To explain why this happens, we consider Scenario I which considers three continuous variables. Although in this case the latent variables correspond to the response, when we use the very vague prior (A) the model tends to over fit the data and thus selecting a different $\bmu_i$ value for each of the 100 data points. The prior choice (B) reduces the number of groups from 90 to around 40, but that's still not enough. So, in order to recover the three true groups, we need to specify a slightly more informative prior as the one given in (C), where the histogram shows a distinctive mode in 3 groups. 
Interestingly, the clustering produced when considering only ordinal variables, like in Scenario II (second row in Figure \ref{fig:nclust}), is less sensible to the choice of the prior on the variances. This is due to the fact that the corresponding latent variables have fixed variance one, so the prior only applies to the variance of the $\{\bmu_i\}$. The mode in the number of groups is 5, 6 and 4 for priors (A), (B) and (C), respectively. Although the mode is in 4 for prior (C), the best clustering, obtained as described in Section \ref{sec:implement}, has 3 groups as desired. 

In Scenario III we have the same two binary variables as in Scenario II plus two more variables, one binary variable that is slightly informative of the clustering and one continuous that is completely unrelated to the underlying clustering. In this scenario the reduction in the number of groups goes from 85 to 6 when moving from prior (A) to (C). Even in the case of prior (C), the model is unable to capture the 3 groups. The best clustering in this case has 5 groups with 3 large groups and two small groups, which is reasonable good. 

\begin{figure}
\centerline{\includegraphics[scale=0.8]{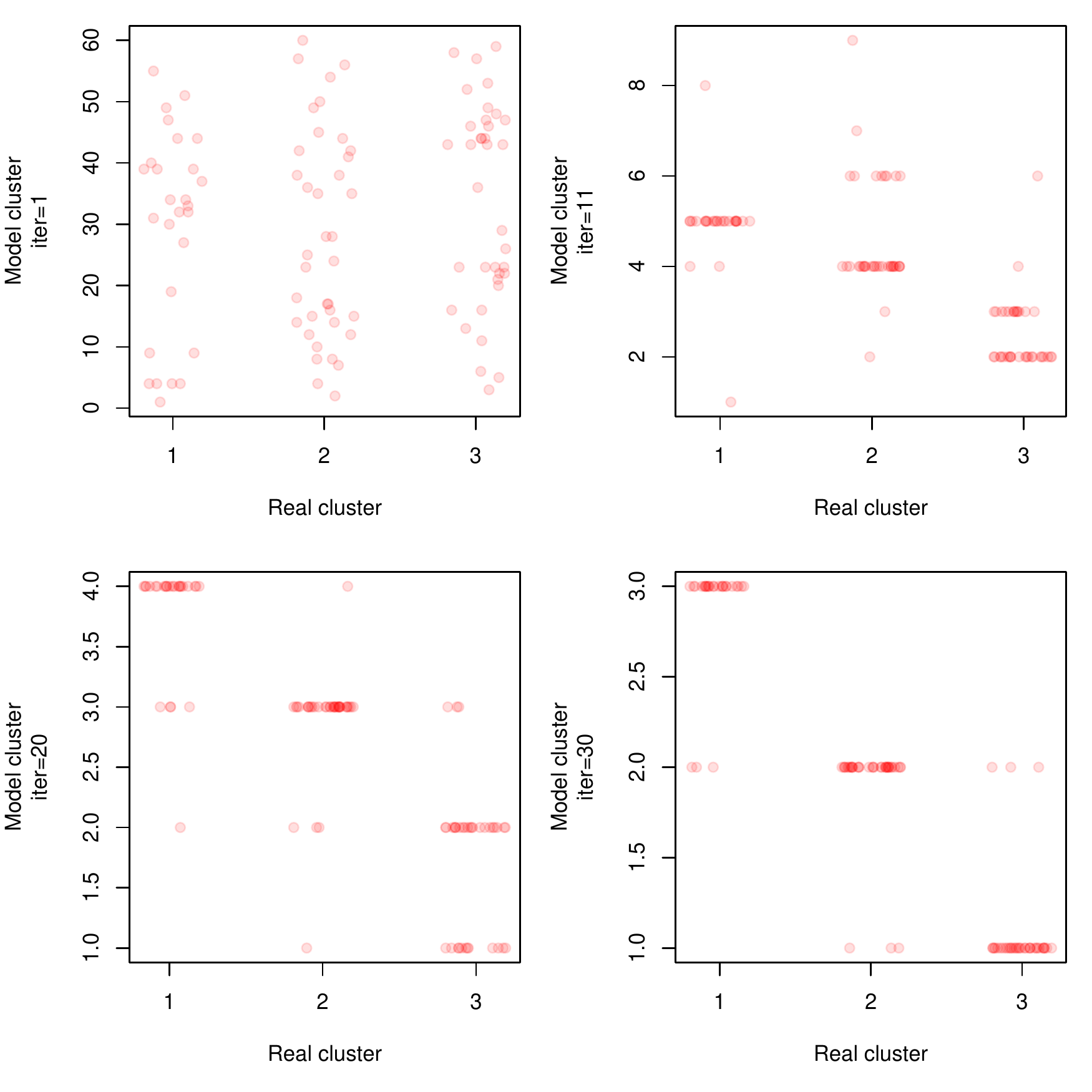}}
\caption{Evolution of the model clustering process at different iterations of the MCMC algorithm for Scenario I(C) in Section \ref{sec:sim1}. Each point corresponds to an individual.}
\label{fig:evolution}
\end{figure}

Finally, we want to show the evolution in the number of groups inferred by the model along the MCMC iterations. Considering Scenario I with prior (C), Figure \ref{fig:evolution} presents the number of groups for iterations 1, 11, 20 and 30 with zero burn-in and no thinning. This graph shows that the speed in detecting the ``correct'' number of clusters is very fast, so there is no need for huge chains if we want to explore the clustering space. 

As mentioned above, no pre-specified values for $a$ and $b$ were taken. These parameters are also very important in the determination of the number of groups. Their estimated values (posterior means) for Scenario I and priors (A), (B) and (C) were 0.99, 0.57 and 0.03 for $a$, and  0.05, 1.29 and 0.52 for $b$, respectively. It is interesting to see that the estimated values for $a$ get smaller, inducing a smaller number of groups, as the prior variances become slightly more informative. 

\subsection{Simulation study 2}
\label{sec:sim2}

\begin{figure}
\centerline{\includegraphics[scale=0.6]{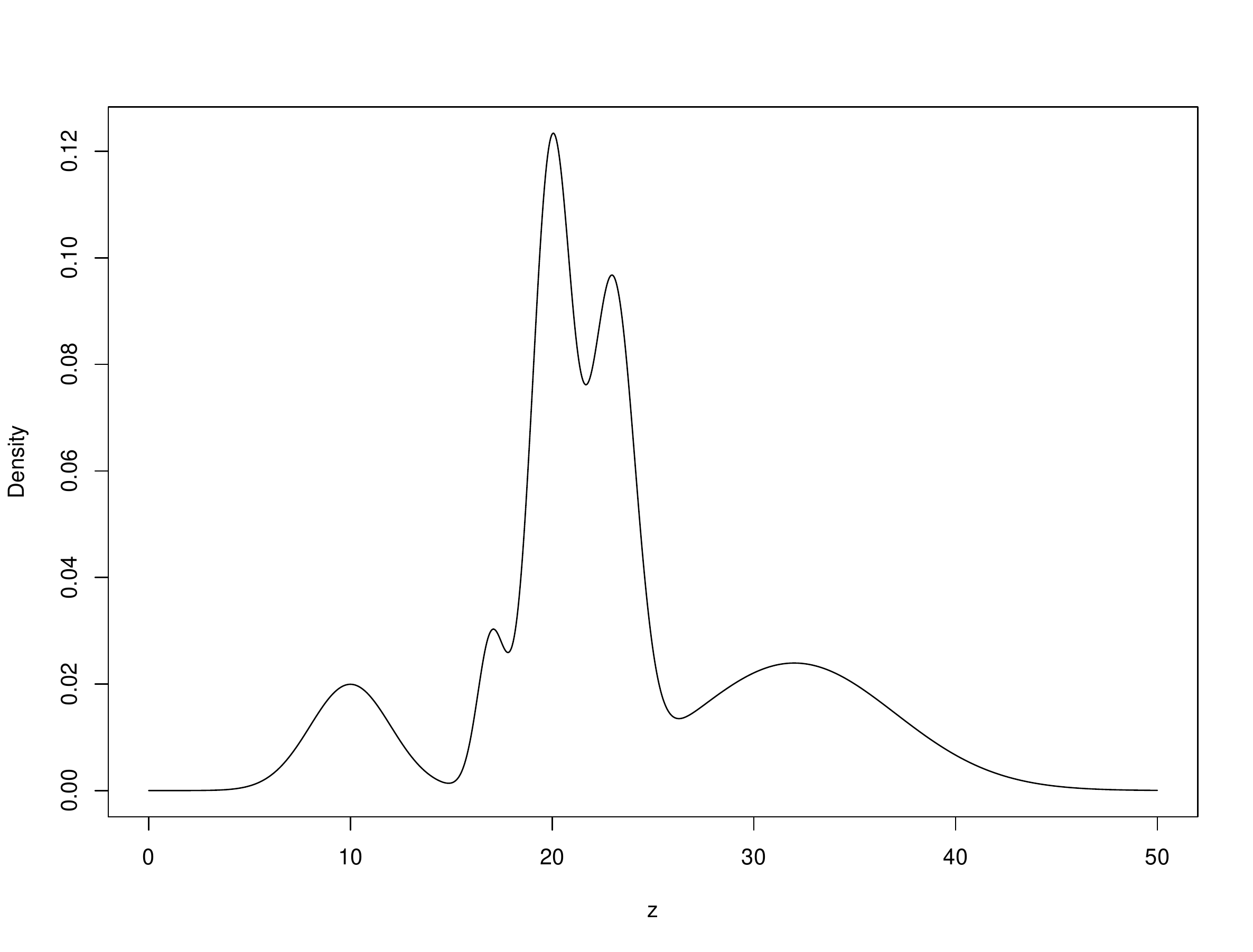}}
\caption{Density used to define sampling probabilities in simulation study of Section \ref{sec:sim2}.} 
\label{fig:sim2}
\end{figure}

In this second simulation study our objective is to show the impact of the sampling probabilities $\pi_i$ and the parameter $\kappa$ in the clustering construction. To illustrate this, consider a univariate density defined by a mixture of five normals with unequal mixing probabilities of the following form: $f(z)=(0.1)\no(z|10,4)+(0.05)\no(z| 17,0.49)+(0.3)\no(z| 20,1)+(0.25)\no(z| 23,1.21)+(0.3)\no(z| 32,25)$. The shape of this density is displayed in Figure \ref{fig:sim2}. It has five modes with the first and the fifth mode well separated, whereas the middle three are kind of overlapped, specially the second mode that is barely distinguished.  

Let us define $n=200$ mutually exclusive intervals $A_i=(\tau_{i-1},\tau_i]$ where $\tau_0=0$ and $\tau_i=\tau_{i-1}+0.25$, for $i=1,\ldots,n$. Calculate $p_i=\P(A_i)$ under density $f(z)$. Simulate a single value $z_i$ uniformly from $A_i$, that is $z_i\sim\un(\tau_{i-1},\tau_i]$, and define $y_i=z_i$, $i=1,\ldots,n$. Clearly, the data $\{y_i\}$ would look like a uniform sample in the interval $(0,50]$. 

The idea is to recover the five groups that exists in a hypothetical population of size $N$ based on the information from this $n=200$ data points and using the probabilities $p_i$ to define the sampling weights $w_i=N p_i$ as in a complex sampling design. Here $\sum_{i=1}^n w_i=N$ and $\bar{w}=N/n$. We will consider three scenarios:
\begin{enumerate}
\item[(IV)] Ignoring the sample design, $\pi_i=1$ and $\kappa=1$
\item[(V)] Acknowledging the sample design, $\pi_i=1/w_i$ and $\kappa=\bar{w}/15$
\item[(VI)] Acknowledging the sample design, $\pi_i=1/w_i$ and $\kappa=\bar{w}/25$,
\end{enumerate}
Note that $w_i$ and $\kappa$ affect multiplicatively the variance of the latent variables, and since $\bar{w}/w_i=\bar{p}/p_i$, where $\bar{p}=(1/n)\sum_{i=1}^n p_i$, there is no need to specify the population size $N$. Scenarios V and VI make the variance in model \eqref{eq:model} to be smaller as compared to the variance induced by Scenario IV and at the same time account for the sampling design. 

We took the prior specifications (C) for the variances and the same prior specifications for the Poisson-Dirichlet process as in the previous simulation study. The Gibbs sampler was ran for 4,700 iterations with a burn in of 200 and keeping one of every 3$^{rd}$ iteration. 

\begin{figure}
\centerline{\includegraphics[scale=0.5]{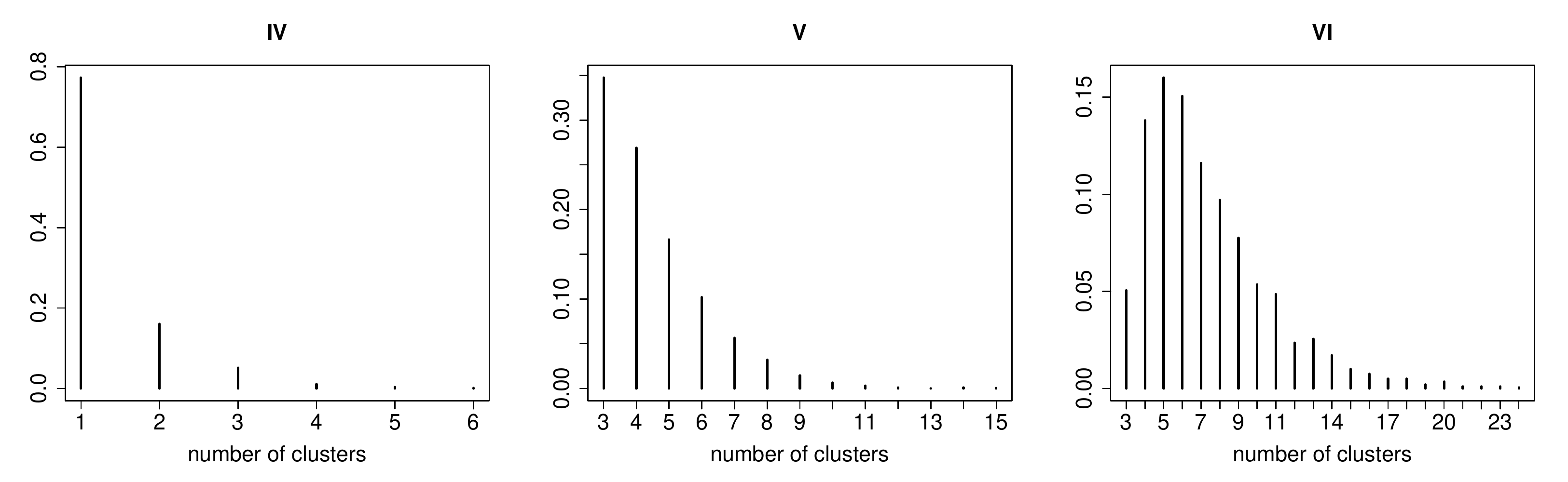}}
\caption{Probability histograms of the number of clusters for scenarios IV, V and V, as described in Section \ref{sec:sim2}.}
\label{fig:nclust2}
\end{figure}

As in the previous simulation study, we summarise the clustering performance by computing the posterior probability of the number of groups obtained by the different runs, reported in Figure \ref{fig:nclust2}.  When we ignore the sample design (first panel), the number of groups is concentrated in a single group around 80\% of the time. For Scenario V, the number of groups is mainly 3, which makes sense since the three middle groups are difficult to distinguish. Finally, the number of groups obtained by Scenario VI has a mode in 5, the correct number of groups.

\subsection{Households data}
\label{sec:house}

\begin{figure}
\centerline{\includegraphics[scale=0.6]{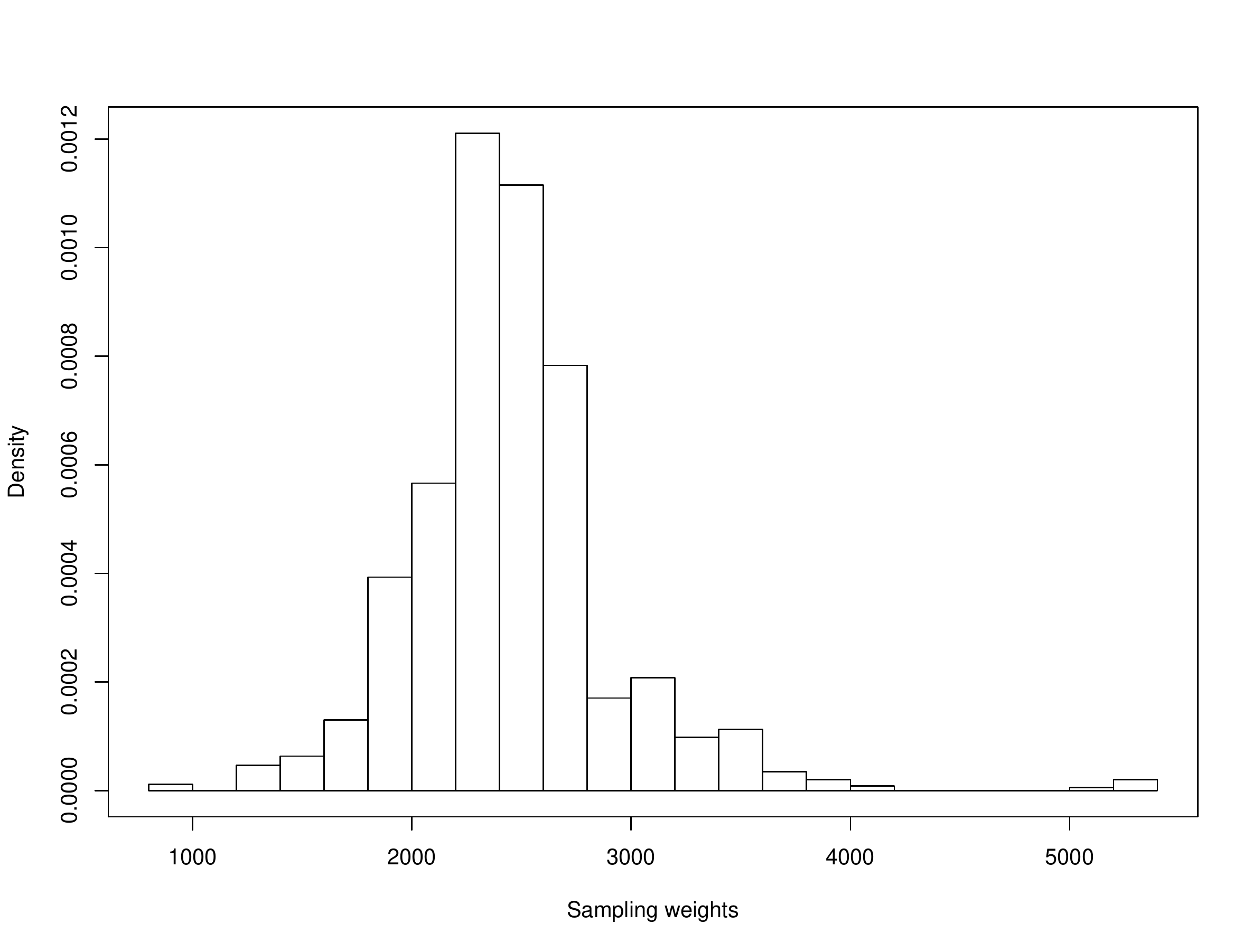}}
\caption{Probability histogram of the sampling design weights or expansion factors $w_i$, for $i=1,\ldots,n$.} 
\label{fig:weights}
\end{figure}

The ENIGH-MCS is an annual nationwide survey in Mexico that is representative to all the 32 states that form the country. For illustration purposes we concentrate in the year 2014 and in one of those states, the ``State of Mexico''. From now on we will refer to it as Edomex for its initials in Spanish. In 2014 this state had $16.2$ million inhabitants, that corresponds to $13.5\%$ of the country population, and had 4.2 million households. Edomex is the most populated state of the country. The survey in Edomex consisted of $n=1,730$ households, which correspond to $0.04\%$ of the total number of households in the state. Since the survey is based on a complex sampling design, each of the households in the sample represents between 960 and 5,286 households with the mode around 2,500 households. A probability histogram of the sampling weights is shown in Figure \ref{fig:weights}.

The analysis will be based on a vector of $p=9$ variables of mixed type. One continuous variable: $Y_1=$ household income (in Mexican pesos); six binary variables: 
$Y_2=$ deprivation to feeding (1--yes, 0--no),
$Y_3=$ deprivation to health services (1--yes, 0--no),
$Y_4=$ housing quality (1--bad, 0--good),
$Y_5=$ education backwardness (1--yes, 0--no), 
$Y_6=$ deprivation to basic public services (1--yes, 0--no),
$Y_7=$ deprivation to social security (1--yes, 0--no);
one ordinal variable: $Y_8=$ education level of the family head (0--incomplete primary, 1--incomplete secondary, 2--complete secondary or more); and one nominal variable: $Y_9=$ town size (1--[100000, $\infty$), 2--[15000, 100000),
3--[2500, 15000), 4--(0, 2500) inhabitants). 

Although variable $Y_9$ can be treated as ordinal, we prefer to treat it as nominal due to the following reasons. The town sizes intervals are unevenly distributed, and moreover, the Ministry of social development believes that poverty conditions are very different in each of these four strata and typically carries out separated analysis for each category.

According to the procedure described in Section \ref{sec:model}, our $p=9$ dimensional observed vector $\bY$ is associated to a latent vector $\bZ$ of dimension $q=11$. This is the result of associating one latent variable to each of the continuous, binary and ordinal variables plus three latent variables for the nominal variable that has four categories. 

For the continuous variable $Y_1$ we require a normalizing transformation to define $Z_1$. An option is to consider the general Box-Cox class with a shift \citep{box&cox:64} and find the best transformation using an optimality criterion. For the case of income variables we suggest to consider a logarithmic transformation with a shift to avoid having problems for the values near zero, that is $Z_1=\log(Y_1+\xi_s)$ with $\xi_s$ the quantile of order $s$ of $Y_1$. Specifically we took $s=0.01$. 
For the binary variables $Y_j$, $j=2,\ldots,7$, the latent variables $Z_j$ have associated thresholds $\gamma_{j,0}=-\infty$, $\gamma_{j,1}=0$ and $\gamma_{j,2}=\infty$, respectively. For the ordinal variable $Y_8$ with $K=3$ ordered categories we define the thresholds as $\gamma_{8,0}=-\infty$, $\gamma_{8,1}=0$, $\gamma_{8,2}=4$ and $\gamma_{8,3}=\infty$. 

The prior specification for the parameters of the Poisson-Dirichlet process were the same as those taken in the simulation study, i.e., $\alpha=0.5$, $d_0^a=d_1^a=d_0^b=d_1^b=1$. To specify the prior on the variances we considered the slightly informative case that arises when $d_0^z=d_0^{\mu}=2.1$ and $d_1^z=d_1^{\mu}=30$. 

For comparison purposes we consider three cases that are described as follows. Let $\bar{w}=(1/n)\sum_{i=1}^n w_i$ then
\begin{enumerate}
\item[i)] Ignoring the sampling design, $\pi_i=1$ and $\kappa=1$
\item[ii)] Acknowledging the sample design, $\pi_i=1/w_i$ and $\kappa=2\bar{w}$
\item[iii)] Acknowledging the sample design, $\pi_i=1/w_i$ and $\kappa=4\bar{w}$
\end{enumerate}

We recall that model \eqref{eq:model} includes a parameter $\kappa$ that acts multiplicative in the variance together with the sampling probabilities $\pi_i$. As was shown in the simulation studies, the variances play a crucial role in determining the clustering structure. Now, since the sampling weights range from $960$ to $5286$, the sampling probabilities range from $0.00019$ to $0.00104$. Therefore, as in Section \ref{sec:sim2}, we get rid of the population size but keep the relative importance of each observation by defining $\kappa$ in terms of $\bar{w}$ as $\kappa=k\bar{w}$, where a high/low value of $k$ induces a smaller/higher number of groups, respectively. 

\begin{figure}
\centerline{
\includegraphics[scale=0.5,clip=true,frame]{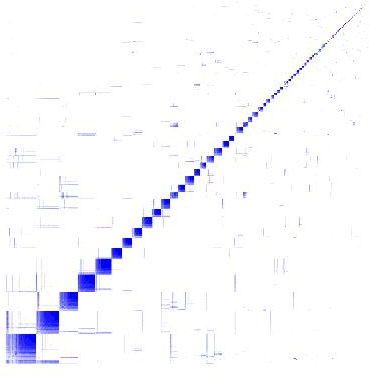}\hspace{0.8cm}
\includegraphics[scale=0.5,clip=true,frame]{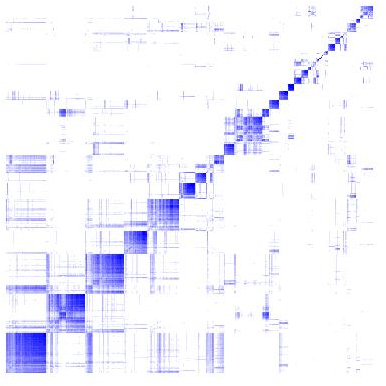} }
\vspace{0.8cm}
\centerline{\includegraphics[scale=0.5,clip=true,frame]{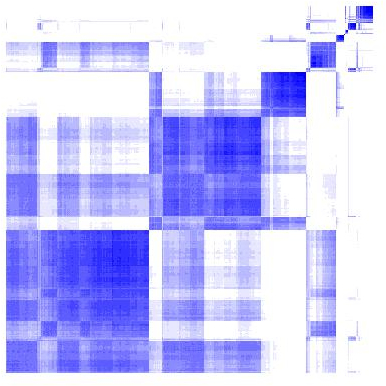}}

\caption{Heatmaps representation of average co-clustering (similarity) matrix for model specifications (i)--(iii) as describen in Section \ref{sec:house}. Ignoring sample weights and $\kappa=1$ (top left), with sample weights and $\kappa=2\bar{w}$ (top right), and $\kappa=4\bar{w}$ (bottom).}
\label{fig:heatmaps}
\end{figure}

The Gibbs sampler was run for 3,200 iterations with a burn in of 200 and a thinning of 3. Results for each of the three cases (i)--(iii) are summarised as heat maps of the average adjacency (clustering) matrix in Figure \ref{fig:heatmaps}. Each graph in the figure correspond to an arranging of the $n=1730$ households and the intensity of the shadows correspond to the estimated probability of these households belonging to the same group. Well formed squares in the inverted diagonal suggest more homogeneous groups. Ignoring the sampling design, case (i), has the consequence of our model producing $r=163$ groups and has a heterogeneity measure \eqref{eq:hm} of $HM=1246$. On the other hand, when considering the sampling design and for the two different values of $\kappa$ we obtain $r=35$ groups  with $HM=2240$ in case (ii) and $r=9$ groups with $HM=3000$ in case (iii). Although the number of groups obtained from case (iii) is smaller, the corresponding heat map and the HM value show that the groups are less homogeneous as compared to the heat maps for cases (i) and (ii). The more homogeneous clustering would be that with 163 groups, however it would be unmanageable. On the other hand, the clustering with 35 groups reduces the number of groups in almost 80\% at a cost of increasing the HM in 79\%. Finally, having only 9 sounds very efficient but the heterogeneity increases in 140\% with respect to the case (i). 

We believe that in applied cluster analysis, the ``best clustering'' does not exist, there are different alternative clusterings and we must choose one according to different criteria including that of a good interpretation of the groups. 

We now proceed to give a brief interpretation of the groups obtained in cases (ii) and (iii). In Tables \ref{tab:clust9} and \ref{tab:clust35} we display the groups (weighted) means for the first eight variables and the (weighted) percentage of households in each of the four categories of the ninth variable. We also include the group size in percentage in the last column. In the last row of the tables we include the population (weighted) means in the whole Edomex for reference. Note that the last value of the table corresponds to the total number of households in Edomex (population size). To simplify the interpretation of the groups we also highlighted in bold those numbers that correspond to a considerably worse poverty condition as compared with the population mean. 

Considering first the clustering of 9 groups, we can say the following. Group 1 is the largest group with 36.0\% of the households in Edomex and its poverty variables show that it is an average group with no particular deprivation different from the state mean. Group 9 has the largest income and the smallest deprivation indicators which make it to be the wealthier group and represents the 0.5\% of the households. The groups 4, 5 and 8 are the poorest groups, with the majority of their values highlighted in bold. Specifically, group 8 has the smallest income and consists of households in rural and semi-rural areas (less than 15000 inhabitants) and represent about 0.7\% of the households in the state. Surprisingly the income of group 6 is not that low, but shows education deficiencies in both education backwardness and education level of the family head. Group 5 presents an interesting feature, it is formed by households from metropolitan areas (more than 100000 inhabitants) and by households from rural areas (less than 2500 inhabitants), but all present feeding, education and public services deprivations.

A clustering with 9 groups is very concise and allows us to identify big chunks of households with certain needs. However a clustering with more groups would allow us to identify households with hidden needs. The clustering of 35 groups might sound too large at a first glance but if we consider that the number of households in the state is around 4 million, then a clustering with 35 groups might not be that large. 

Analysing Table \ref{tab:clust35}, more specific needs can be detected in the groups. For instance group 15 consists of households in urban areas (more than 15000 inhabitants) but have low income, feeding problems and no access to social security. This group is hidden in the thicker clustering of Table \ref{tab:clust9}. Group 28, on the other hand, has the second largest income, people live in metropolitan and in semi-urban areas, but show problems related to access to education. Additionally, group 2 has an average income with households in metropolitan areas and shows a lack of access to social security. 
Another important aspect of this thinner clustering is that there are no large groups, the group with most households, group 1, has 11.78\% of the total, whereas the smallest group, group 35, has 0.05\% of the households. This is in contrast to the clustering with 9 groups where almost 90\% of the households concentrate in 4 groups.

\section{Concluding remarks}
\label{sec:conclusion}

Multivariate cluster analysis is one of the most useful techniques in practice and there are many methods available in all statistical software. Most of these techniques typically require the data to be numeric and preferably continuous. However, when the multivariate data contains mixed mode variables like continuous, discrete, ordinal and nominal, the clustering task cannot be performed straightforwardly.

We discussed a nonparametric model-based clustering approach that is entirely flexible and able to perform clustering of individuals with multivariate mixed-scale variables. The model relies on the introduction of latent continuous variables modelled via a multivariate mixture model. The mixing distribution is kept unspecified and totally flexible assuming a nonparametric Poisson-Dirichlet process prior. Its almost sure discreteness naturally accounts for the desired clustering structure. 

Our model allows for the treatment of data coming from a complex design, as in our motivating application about the Mexican households survey, including the sampling weights in the analysis. 

Our findings suggest the importance of the mixture component variances in the clustering procedure and specifically the lower the variances, the higher the number of groups and viceversa. Thus, according to the desired needs, the prior distributions for the variances and any scaling factor need to be carefully elicited. As long as the continuous data are standardized to have unit variance, a vague inverse gamma distribution with shape parameter close to 2 and high scale parameter (above 10) could be used as default.

To report a final clustering of the observations, we proposed to report the clustering of all MCMC iterations, with smallest squared distance to the average co-clustering (similarity) matrix, as described in Section \ref{sec:implement}. Alternatively, the HM measure \eqref{eq:hm} can be computed for the clusterings of all MCMC iterations and the clustering with the smallest HM can also be reported. The current version of the R-package {\tt BNPMIXcluster} implements this. 

As for the particular application, our clustering tool will be very beneficial for the authorities in charge of creating social programmes as it allows to identify those households with specific needs and to see the potential number of households that would benefit with the implementation of a social program. Moreover, if so desired, it is also possible to include geographical coordinates as longitude and latitude as part of the measured variables to produce spatially cohesive clusters. 

Our model is based on a location mixtures of normals. Extensions of our model to more general kind of mixtures are also possible. For instance a location-scale mixture would produce different variance-covariance matrices $\bSigma$ for each group. This would be more computationally intensive, but might be worth exploring.

\section*{Acknowledgements}
The authors are grateful to the constructive comments of a guest editor and two anonymous referees. 
The second author acknowledges support from \textit{Asociaci\'on Mexicana de Cultura, A. C.} Mexico. 
The third author is also affiliated with the Collegio Carlo Alberto and acknowledges support of grant CPDA154381/15 from the University of Padova, Italy.

\bibliographystyle{natbib}

\newpage

\begin{table}
\caption{Group means for the clustering with 9 groups obtained with case (iii). Columns are divided according to the nature of the variables: continuous, binary, ordinal and nominal. Last row is the population mean, apart from the last value that corresponds to the population size.}
\label{tab:clust9}
{\scriptsize
\begin{center}
\begin{tabular}{c|r|cccccc|c|cccc|r} \hline \hline
group	&	income	&	feed	&	health	&	house	&	edu	&	serv	&	ss	&	hedu	&	ts:1	&	ts:2	&	ts:3	&	ts:4	&	size	\\ \hline
1	&	5934	&	0.22	&	0.35	&	0.11	&	0.32	&	0.11	&	0.80	&	1.91	&	0.59	&	0.13	&	0.15	&	0.13	&	36.0\%	\\
2	&	11374	&	0.16	&	0.34	&	0.07	&	0.36	&	0.07	&	0.71	&	1.93	&	0.64	&	0.12	&	0.14	&	0.10	&	30.5\%	\\
3	&	22682	&	0.06	&	0.34	&	0.01	&	0.23	&	0.04	&	0.67	&	1.99	&	0.74	&	0.09	&	0.10	&	0.06	&	12.4\%	\\
4	&	{\bf 3091}	&	0.26	&	{\bf 0.42}	&	0.06	&	0.43	&	{\bf 0.24}	&	{\bf 0.89}	&	1.78	&	0.54	&	0.11	&	0.16	&	0.20	&	9.0\%	\\
5	&	{\bf 1783}	&	{\bf 0.37}	&	0.23	&	0.09	&	{\bf 0.95}	&	{\bf 0.41}	&	0.60	&	{\bf 0.21}	&	{\bf 0.41}	&	0.05	&	0.18	&	{\bf 0.36}	&	4.0\%	\\
6	&	5006	&	0.25	&	0.32	&	0.13	&	{\bf 0.94}	&	0.15	&	0.64	&	{\bf 0.53}	&	0.46	&	0.11	&	0.26	&	0.16	&	4.2\%	\\
7	&	44991	&	0.04	&	0.09	&	0.00	&	0.14	&	0.02	&	0.30	&	2.00	&	0.63	&	0.29	&	0.06	&	0.02	&	2.7\%	\\
8	&	{\bf 570}	&	{\bf 0.69}	&	0.24	&	{\bf 0.46}	&	{\bf 0.68}	&	{\bf 0.29}	&	{\bf 1.00}	&	1.46	&	0.16	&	0.06	&	{\bf 0.48}	&	{\bf 0.30}	&	0.7\%	\\
9	&	219578	&	0.16	&	0.27	&	0.00	&	0.00	&	0.00	&	0.27	&	2.00	&	0.89	&	0.11	&	0.00	&	0.00	&	0.5\%	\\ \hline
pop	&	11212	&	0.19	&	0.34	&	0.08	&	0.38	&	0.11	&	0.74	&	1.79	&	0.61	&	0.12	&	0.15	&	0.13	&	4240837	\\
\hline \hline
\end{tabular}
\end{center}
}
\end{table}

\begin{table}
\caption{Group means for the clustering with 44 groups obtained with case (ii). Columns are divided according to the nature of the variables: continuous, binary, ordinal and nominal. Last row is the population mean, apart from the last value that corresponds to the population size.}
\label{tab:clust35}
{\scriptsize
\begin{center}
\begin{tabular}{c|r|cccccc|c|cccc|r} \hline \hline
group	&	income	&	feed	&	health	&	house	&	edu	&	serv	&	ss	&	hedu	&	ts:1	&	ts:2	&	ts:3	&	ts:4	&	size	\\ \hline
1	&	6305	&	0.08	&	0.08	&	0.07	&	0.17	&	0.13	&	{\bf 0.90}	&	1.97	&	0.56	&	0.14	&	0.14	&	0.17	&	11.78\%	\\
2	&	11386	&	0.08	&	0.39	&	0.06	&	0.04	&	0.01	&	{\bf 0.95}	&	1.97	&	0.69	&	0.11	&	0.13	&	0.07	&	8.87\%	\\
3	&	6106	&	0.16	&	{\bf 0.89}	&	0.02	&	0.19	&	0.03	&	{\bf 0.96}	&	1.97	&	0.74	&	0.10	&	0.09	&	0.07	&	7.57\%	\\
4	&	9126	&	0.11	&	0.02	&	0.05	&	0.14	&	0.03	&	0.08	&	1.97	&	0.74	&	0.12	&	0.09	&	0.05	&	7.02\%	\\
5	&	23174	&	0.04	&	0.06	&	0.01	&	0.13	&	0.02	&	0.58	&	2.00	&	{\bf 0.83}	&	0.06	&	0.08	&	0.04	&	6.54\%	\\
6	&	8744	&	0.06	&	{\bf 0.71}	&	0.22	&	{\bf 0.86}	&	0.18	&	{\bf 0.96}	&	1.94	&	0.58	&	0.09	&	0.22	&	0.11	&	5.77\%	\\
7	&	3883	&	0.20	&	0.59	&	0.07	&	{\bf 0.79}	&	{\bf 0.30}	&	{\bf 0.99}	&	1.90	&	0.31	&	0.14	&	0.23	&	0.31	&	5.64\%	\\
8	&	14468	&	0.07	&	0.04	&	0.02	&	0.10	&	0.05	&	0.07	&	2.00	&	0.64	&	0.18	&	0.13	&	0.05	&	4.47\%	\\
9	&	2613	&	0.08	&	0.31	&	0.07	&	0.14	&	0.19	&	{\bf 0.95}	&	1.94	&	0.54	&	0.18	&	0.10	&	0.18	&	3.95\%	\\
10	&	5991	&	{\bf 0.91}	&	0.10	&	0.09	&	0.24	&	0.11	&	0.82	&	1.98	&	0.50	&	0.19	&	0.21	&	0.11	&	3.33\%	\\
11	&	21084	&	0.08	&	{\bf 0.83}	&	0.00	&	0.04	&	0.08	&	{\bf 0.96}	&	1.98	&	0.66	&	0.11	&	0.13	&	0.10	&	3.53\%	\\
12	&	19577	&	0.12	&	{\bf 0.83}	&	0.06	&	{\bf 0.94}	&	0.05	&	{\bf 0.95}	&	1.97	&	0.65	&	0.11	&	0.19	&	0.04	&	3.14\%	\\
13	&	12034	&	0.13	&	0.04	&	0.08	&	{\bf 0.86}	&	0.05	&	{\bf 0.93}	&	1.97	&	0.59	&	0.11	&	0.13	&	0.17	&	3.19\%	\\
14	&	4434	&	0.02	&	0.00	&	0.06	&	0.14	&	0.02	&	0.02	&	1.83	&	0.75	&	0.14	&	0.02	&	0.09	&	2.79\%	\\
15	&	3421	&	{\bf 0.87}	&	0.43	&	0.06	&	0.34	&	0.07	&	{\bf 0.97}	&	1.95	&	{\bf 0.77}	&	0.02	&	0.17	&	0.04	&	2.58\%	\\
16	&	10546	&	{\bf 0.88}	&	{\bf 0.73}	&	0.12	&	{\bf 0.72}	&	{\bf 0.25}	&	{\bf 0.98}	&	1.89	&	0.59	&	0.11	&	0.14	&	0.16	&	2.48\%	\\
17	&	3027	&	0.02	&	0.10	&	0.07	&	{\bf 0.94}	&	0.16	&	0.36	&	{\bf 0.41}	&	0.50	&	0.02	&	0.25	&	0.23	&	2.40\%	\\
18	&	7541	&	0.13	&	0.14	&	0.00	&	{\bf 0.87}	&	0.18	&	0.68	&	1.05	&	0.60	&	0.04	&	0.21	&	0.15	&	2.25\%	\\
19	&	41828	&	0.00	&	0.02	&	0.00	&	0.02	&	0.00	&	0.18	&	2.00	&	0.61	&	{\bf 0.33}	&	0.01	&	0.04	&	2.68\%	\\
20	&	{\bf 1765}	&	0.44	&	0.09	&	0.08	&	{\bf 0.97}	&	{\bf 0.74}	&	0.60	&	{\bf 0.08}	&	0.15	&	0.11	&	0.14	&	{\bf 0.60}	&	1.83\%	\\
21	&	4645	&	0.64	&	0.22	&	{\bf 0.35}	&	{\bf 0.91}	&	{\bf 0.37}	&	{\bf 0.95}	&	{\bf 0.88}	&	0.27	&	0.15	&	0.27	&	0.32	&	1.77\%	\\
22	&	5272	&	0.26	&	{\bf 0.78}	&	0.07	&	{\bf 0.93}	&	0.04	&	0.85	&	{\bf 0.36}	&	{\bf 0.48}	&	0.08	&	{\bf 0.36}	&	0.08	&	1.53\%	\\
23	&	4219	&	0.15	&	0.44	&	{\bf 0.86}	&	0.15	&	0.10	&	0.80	&	1.94	&	0.65	&	0.16	&	0.20	&	0.00	&	1.12\%	\\
24	&	{\bf 1406}	&	{\bf 0.89}	&	0.43	&	0.28	&	{\bf 0.80}	&	0.10	&	0.88	&	{\bf 0.78}	&	0.61	&	0.00	&	0.23	&	0.16	&	0.96\%	\\
25	&	{\bf 1518}	&	0.11	&	{\bf 0.83}	&	0.00	&	{\bf 1.00}	&	0.00	&	{\bf 1.00}	&	{\bf 0.25}	&	0.35	&	0.00	&	0.51	&	0.14	&	0.42\%	\\
26	&	{\bf 730}	&	{\bf 0.86}	&	0.42	&	0.28	&	0.69	&	0.13	&	{\bf 1.00}	&	1.84	&	0.42	&	0.00	&	0.58	&	0.00	&	0.39\%	\\
27	&	{\bf 385}	&	0.27	&	0.26	&	0.28	&	{\bf 1.00}	&	{\bf 0.88}	&	{\bf 1.00}	&	{\bf 0.25}	&	0.12	&	0.19	&	0.15	&	{\bf 0.54}	&	0.39\%	\\
28	&	55861	&	0.25	&	0.41	&	0.00	&	{\bf 0.84}	&	0.13	&	0.69	&	2.00	&	0.75	&	0.00	&	0.25	&	0.00	&	0.44\%	\\
29	&	219578	&	0.16	&	0.27	&	0.00	&	0.00	&	0.00	&	0.27	&	2.00	&	{\bf 0.89}	&	0.11	&	0.00	&	0.00	&	0.46\%	\\
30	&	11588	&	0.17	&	0.22	&	0.00	&	{\bf 1.00}	&	0.00	&	0.22	&	{\bf 0.00}	&	0.56	&	0.22	&	0.00	&	0.22	&	0.27\%	\\
31	&	4678	&	0.34	&	0.00	&	{\bf 0.66}	&	{\bf 1.00}	&	{\bf 0.66}	&	0.00	&	{\bf 0.34}	&	0.34	&	0.00	&	0.34	&	0.32	&	0.15\%	\\
32	&	{\bf 225}	&	0.45	&	0.00	&	0.00	&	0.00	&	0.00	&	{\bf 1.00}	&	2.00	&	0.00	&	{\bf 0.45}	&	0.00	&	{\bf 0.55}	&	0.10\%	\\
33	&	9298	&	{\bf 1.00}	&	{\bf 1.00}	&	{\bf 1.00}	&	{\bf 1.00}	&	0.00	&	{\bf 1.00}	&	2.00	&	{\bf 1.00}	&	0.00	&	0.00	&	0.00	&	0.05\%	\\
34	&	18055	&	{\bf 1.00}	&	0.00	&	0.00	&	{\bf 1.00}	&	0.00	&	0.00	&	1.00	&	{\bf 1.00}	&	0.00	&	0.00	&	0.00	&	0.06\%	\\
35	&	{\bf 2376}	&	{\bf 1.00}	&	{\bf 1.00}	&	{\bf 1.00}	&	{\bf 1.00}	&	0.00	&	{\bf 1.00}	&	2.00	&	0.00	&	{\bf 1.00}	&	0.00	&	0.00	&	0.05\%	\\ \hline
pop	&	11212	&	0.19	&	0.34	&	0.08	&	0.38	&	0.11	&	0.74	&	1.79	&	0.61	&	0.12	&	0.15	&	0.13	&	4240837	\\
\hline \hline
\end{tabular}
\end{center}
}
\end{table}

\end{document}